\documentclass{an}
\usepackage{graphicx}
\usepackage{times}
\usepackage{fancyhdr}
\usepackage{natbib}
\begin{document}
\include{inc_header}

\title{QPOs in microquasars and Sgr$\,$A${^*}$: measuring the black hole spin
}

\author{G. T{\"o}r{\"o}k\inst{1,2}}

\institute{
Institute of Physics, Faculty of Philosophy and Science, Silesian University in Opava,
Bezru{\v c}ovo n\'am. 13, 746 01 Opava, Czech Republic
   \and
Nordita, Blegdamsvej 17, 2100 Copenhagen, Denmark	}
       
\date{Receivedv 8 September 2005; accepted 20 September 2005;
published online 20 October 2005}

\abstract{
In all four microquasars which show double peak kHz QPOs, the ratio of the two frequencies is 3:2. This strongly supports the suggestion that twin peak kHz QPOs are due to a resonance between some modes of accretion disk oscillations. Here, we stress that fits to observations of the hypothetical resonances between vertical and radial epicyclic frequencies (particularly of the parametric resonance) give an accurate estimate of the spin for the three microquasars with known mass. Measurement of double peak QPOs frequencies in the Galaxy centre seems also to be consistent with the 3:2 ratio established by previous observations in microquasars, however the Sgr${\,}$A$^*$ data are rather difficult for the same exact analysis. If confirmed, the 3:2 ratio of double peak QPOs in Sgr$\,$A$^*$ would be of a fundamental importance for the black hole accretion theory and the precise measurement could help to solve the question of QPOs nature.
}

\correspondence{terek@volny.cz}

\maketitle

\section{Orbital resonance models -- the brief sketch}

\citet{KluzniakAbramowicz2000}, motivated by the observed double peak kHz QPOs in X--ray variability of neutron stars sources, introduced models based on the idea of a resonance between the relativistic epicyclic frequencies
\footnote{Before the first discovery of the double peak QPO in black hole source GRO 1655--40 by \citet{Strohmayer2001}, \citet{KluzniakAbramowicz2000} suggested on theoretical ground that such eventual QPOs should have rational ratios.}
.
These models were reviewed, e.g., in \citet{AbramowiczKluzniak2004,TAKS} or \citet{KluzniakNordita}. 

According to the resonance hypothesis, the two modes in resonance have eigenfrequencies $\nu_{\rm rad}$ (equal to the radial epicyclic frequency) and $\nu_{\rm v}$ (equal to the vertical orbital frequency $\nu_{\theta}$ or to the Keplerian frequency $\nu_{\rm K}$). Several resonances of this kind are possible, and have been  discussed (see, e.g., \citealt[]{AbramowiczKluzniak2004}).
 Main relations between the eigenfrequencies of resonance and the possibly observable frequencies are summarized in Table \ref{table1} for some particular resonances.

\begin{table}[h]
\begin{center}
      \caption[]{\label{table1}Relation of observed frequencies for standard~($\nu=\nu_\theta$) and ``Keplerian''~($\nu=\nu_\mathrm{K}$) resonances -- while the parametric model identifies the observed $\nu_\mathrm{up}$,~$\nu_\mathrm{down}$ directly with the eigenfrequencies of resonance, forced resonances can give the 3:2 ratio by combinational frequencies.}
\renewcommand{\arraystretch}{1.2}
\begin{tabular}{ c  c  c  c  c  c }
     \hline
    \multicolumn{2}{c}{Type of resonance}&&&
     \multicolumn{2}{c}{Observed frequencies}\\ 
    \multicolumn{2}{c}{($\mathrm{n}\nu_{\mathrm{rad}}=\mathrm{m}\nu_\mathrm{}$)}&~~~n~~~&~~~m~~~&
      $~~~\nu_\mathrm{upp}~~~~$ &$~~~\nu_\mathrm{down}~~~$ \\
     \hline
     \hline
     &parametric&3&2&$\nu_\theta$&$\nu_\mathrm{rad}$\\
     \cline{2-6}
   \rotatebox{90}{\makebox(0,0){~~standard}} 
   &3:1 forced&3&1&$\nu_\theta$&$\nu_\theta-\nu_\mathrm{rad}$\\
    \cline{2-6}
    &2:1
    forced&2&1&$\nu_\theta+\nu_\mathrm{rad}$&$\nu_\theta$\\
    \hline

     \hline
     &parametric&3&2&$\nu_\mathrm{K}$&$\nu_\mathrm{rad}$\\
     \cline{2-6}
   \rotatebox{90}{\makebox(0,0){~~Keplerian}} 
   &3:1 forced&3&1&$\nu_\mathrm{K}$&$\nu_\mathrm{K}-\nu_\mathrm{rad}$\\
    \cline{2-6}
    &2:1
    forced&2&1&$\nu_\mathrm{K}+\nu_\mathrm{rad}$&$\nu_\mathrm{K}$\\
    \hline
    \end{tabular} 
 \end{center}
\end{table}


\begin{figure*}[t!]
\centering
\begin{minipage} {.47\hsize}
\includegraphics[width=\hsize]{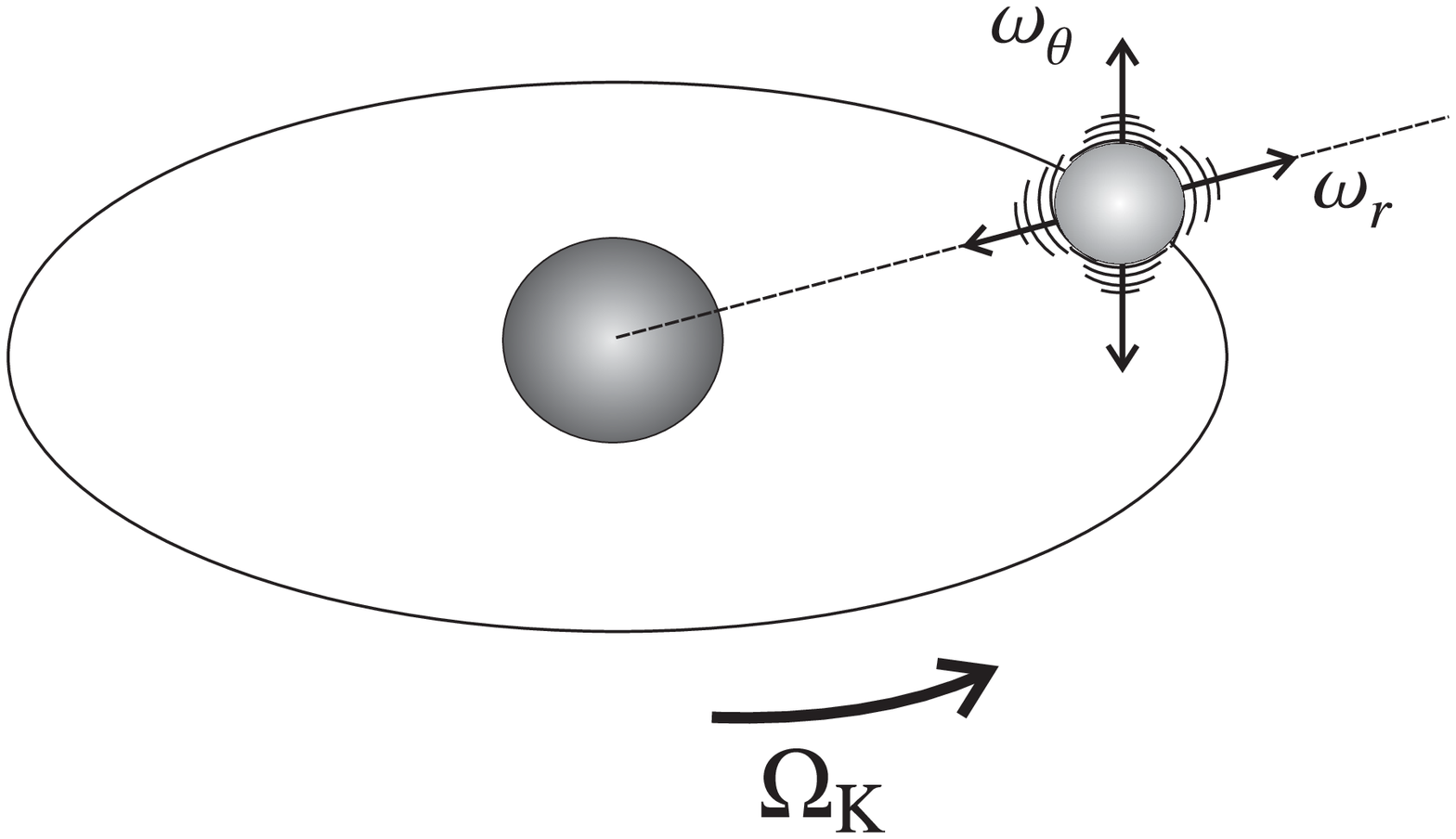}
\end{minipage}
\hfill
\begin{minipage} {.48\hsize}
\includegraphics[angle=-90, width=\hsize]{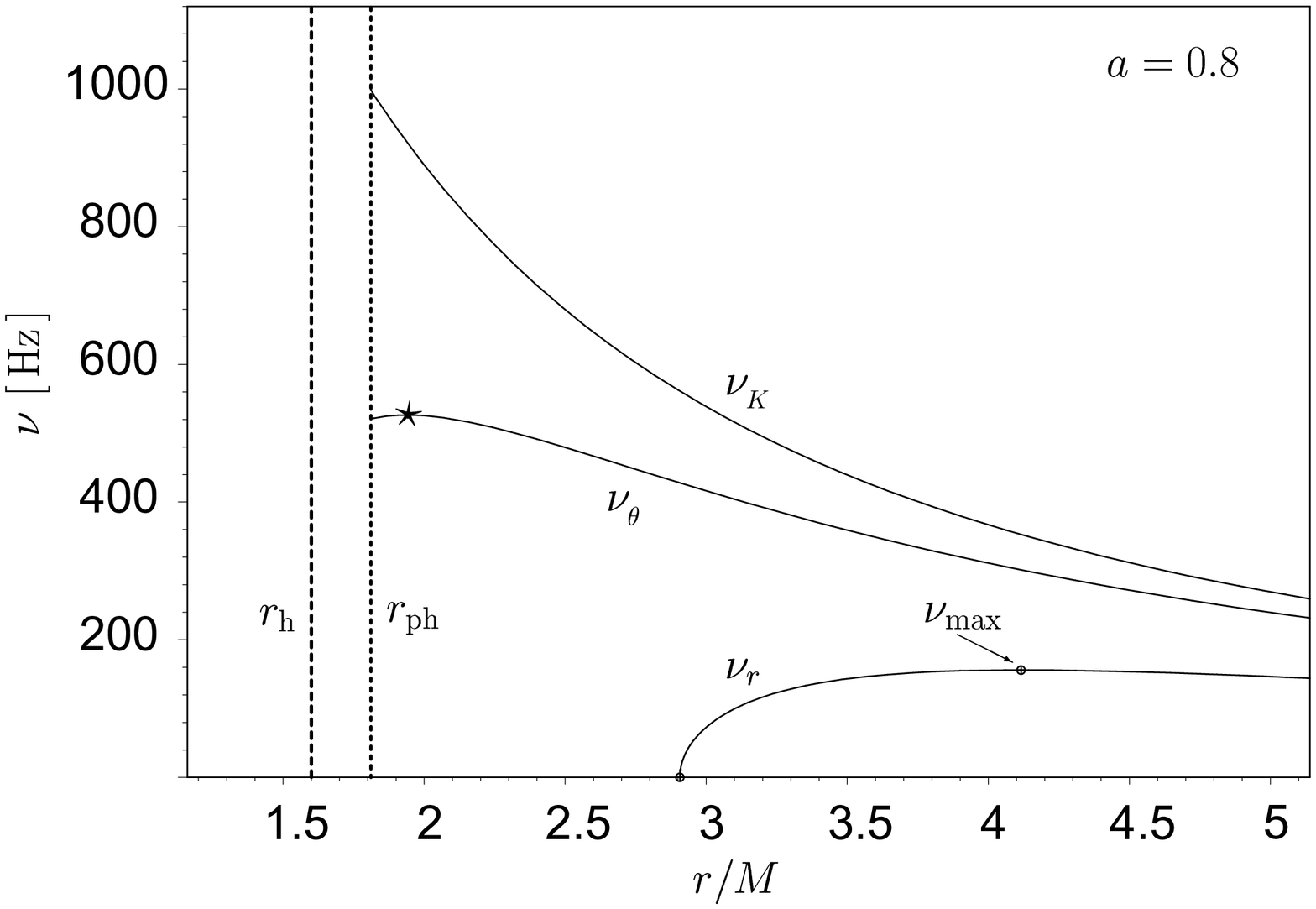}
\end{minipage}
\caption{\label{Figure1} {\it Left cartoon:} In the gravitational field of a central object, test particle on a circular orbit starts oscillate after the small perturbation. Frequencies of these oscillations (radial $\nu_\mathrm{r}$ and vertical $\nu_\theta$) are fundamentally different in Newton's and Einstein's gravity. In Newtonian physics these epicyclic frequencies must always be equal to the Keplerian frequency of circular orbit and the resulting trajectory is an ellipse, while  in Einstein's theory they differ and trajectory is not closed.
{\it Right:} Figure plotted for moderately rotating $10\,{\rm M}_\odot$ black hole shows behaviour of epicyclic frequencies typical for Kerr black holes -- strong Einstein's gravity makes $\nu_{\rm K}\geq\nu_\theta>\nu_r$.
}
\end{figure*}

\begin{figure*}[t!]
\centering
\begin{minipage} {.32\hsize}

\includegraphics[angle=0, width=\hsize]{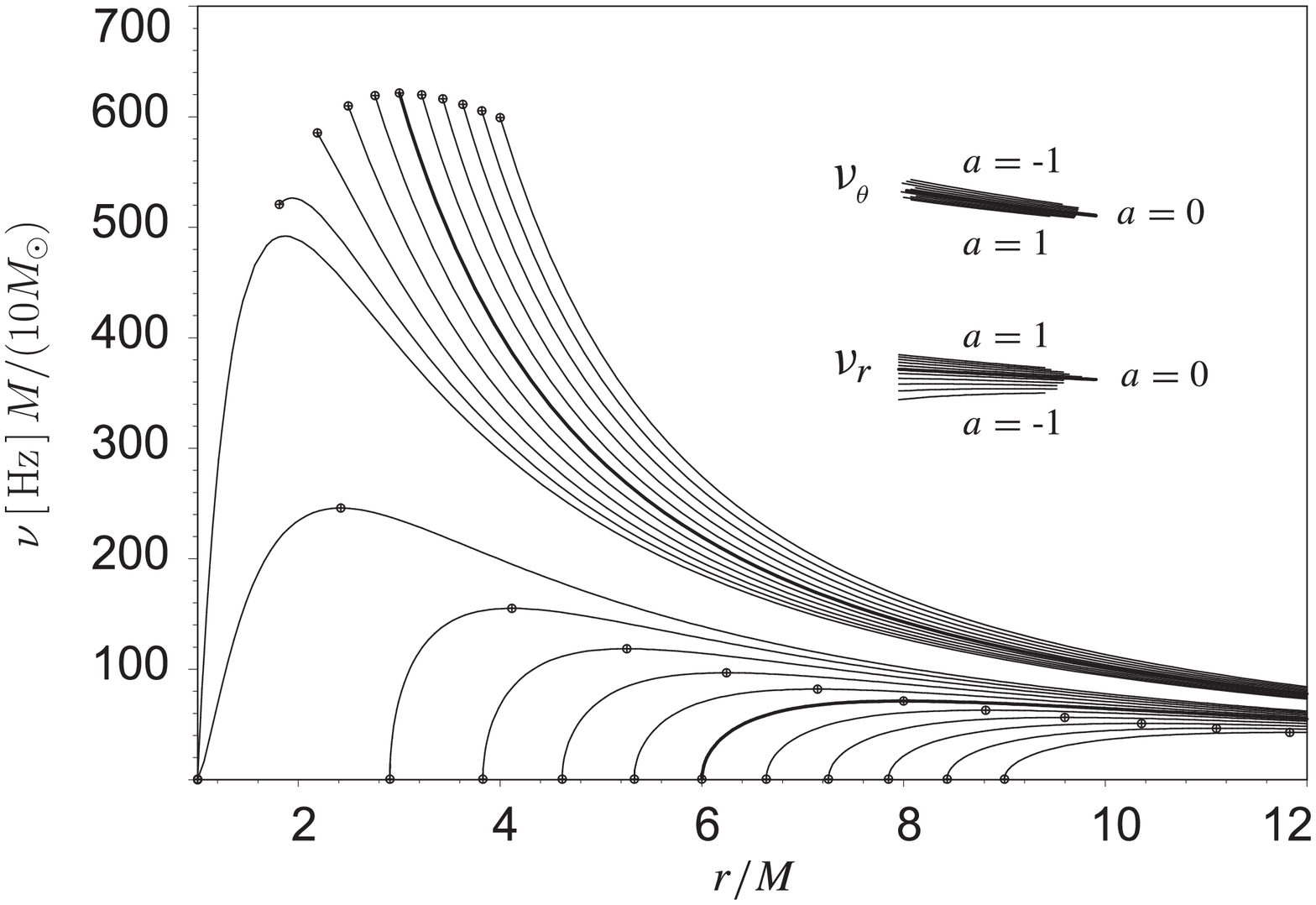}
\end{minipage}
\hfill
\begin{minipage} {.32\hsize}
\includegraphics[angle=-90, width=\hsize]{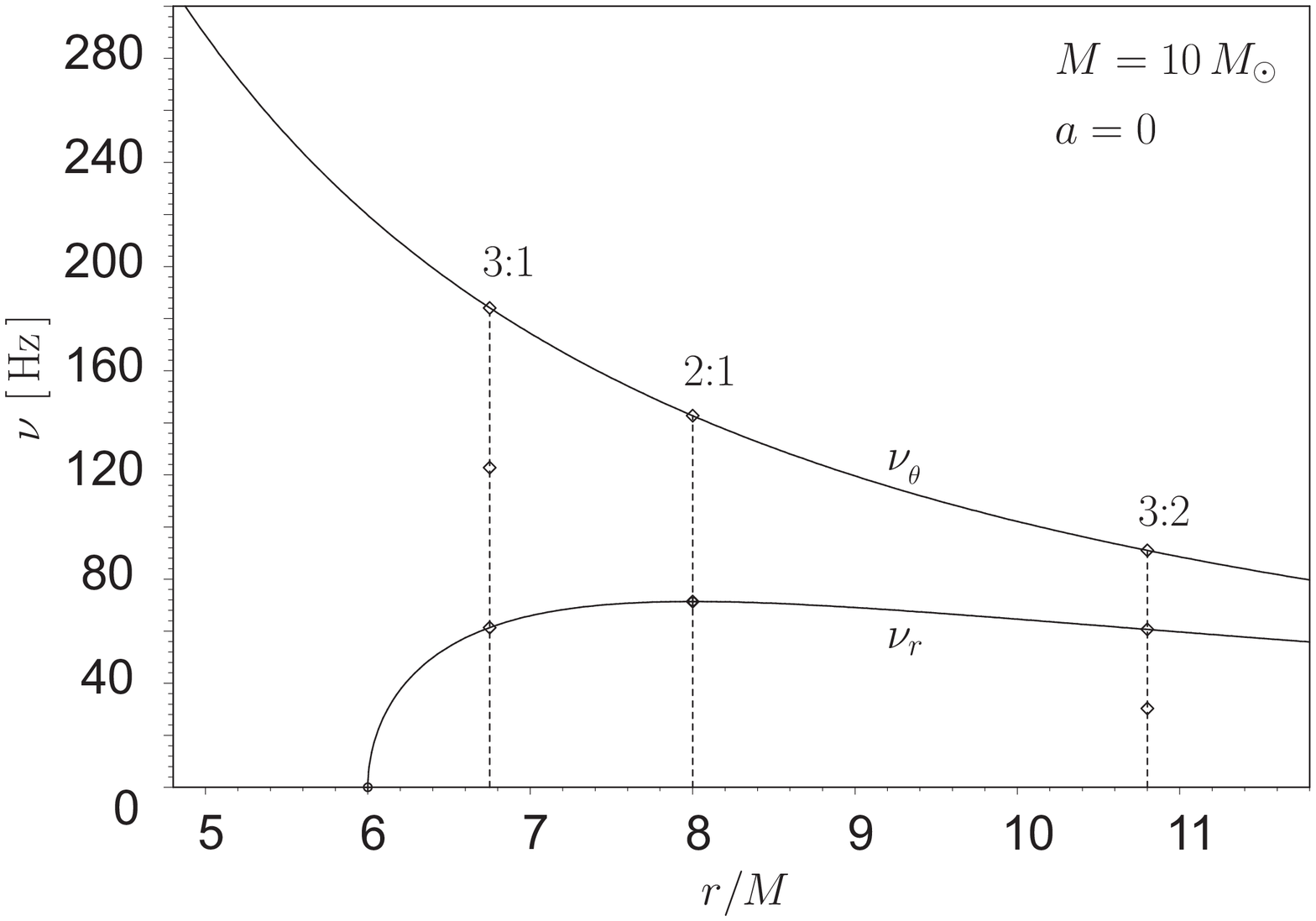}
\end{minipage}
\hfill
\begin{minipage} {.32\hsize}
\includegraphics[angle=0, width=\hsize]{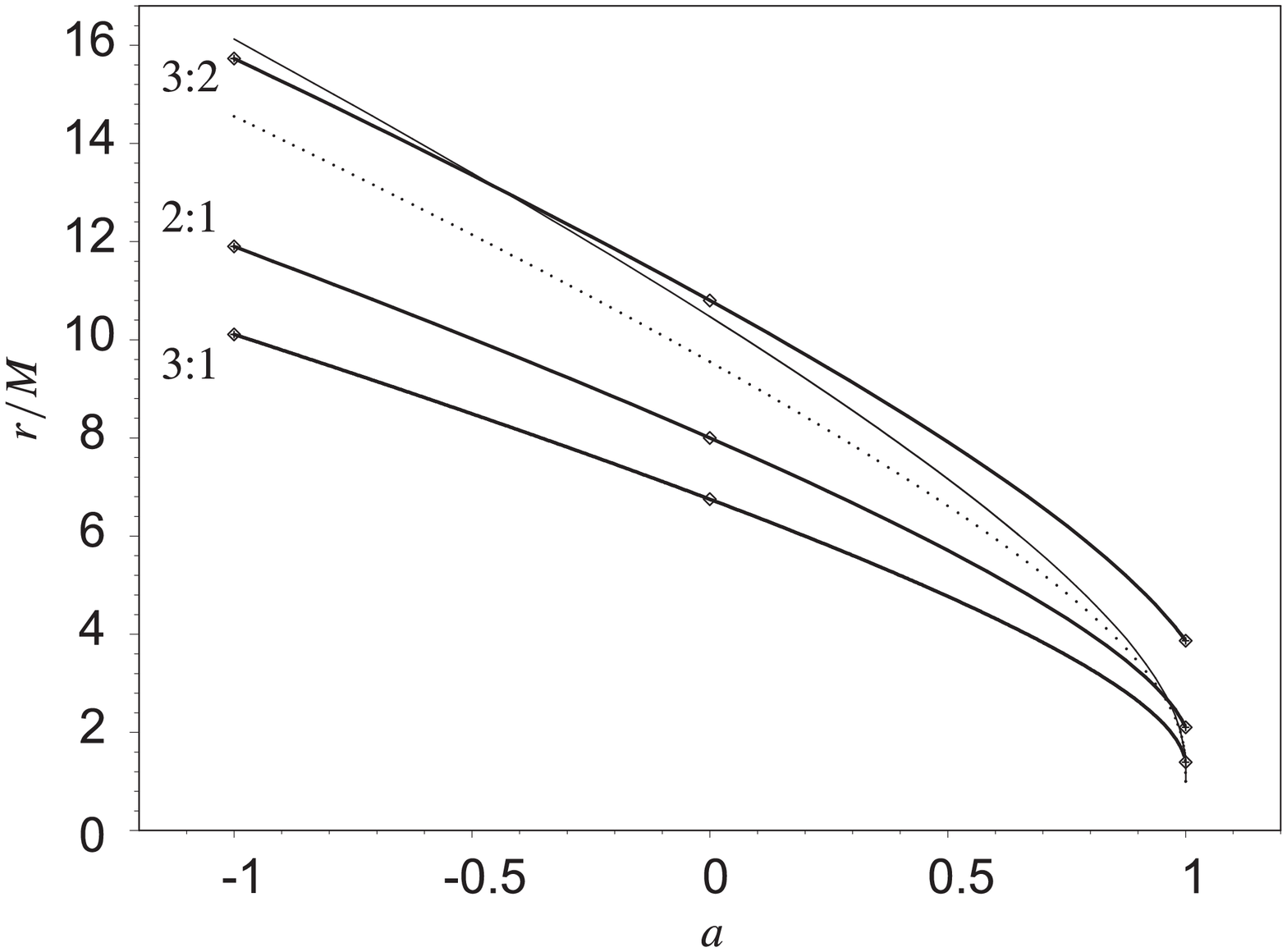}
\end{minipage}
\caption{\label{Figure2}{\it Left:} Behaviour of epicyclic frequencies is sensitiv to the spin. Here, the curves are spaced every 0.2 in $a$.
{\it Middle:} Locations of the three particular resonances for the Schwarzschild black hole ($a=0$).
{\it Right:} Locations of the same three resonances as functions of the black hole spin (discontinuos lines indicate the position of main part of the accretion disc for two idealized cases of thin and thick disc).
}
\end{figure*}

\section{Estimating the black hole spin}

Formulae for the Keplerian and epicyclic frequencies $\nu_{\rm vert}$ and $\nu_{\rm rad}$ in the gravitational field of a rotating Kerr black hole with the mass $M$ and internal angular momentum $a$ (here and henceforth \emph{spin} $a$) are well known (Fig. \ref{Figure1}) -- see, e.g., \citet{NowakLehr1999}:
\newpage

\begin{eqnarray}
\label{frequencies}
\nu_{\mathrm{K}} & = & {1\over 2\pi}\left ({{GM_0}\over {r_G^{~3}}}\right )^{1/2}\left( x^{3/2} + a \right)^{-1},
\nonumber\\
\nu_{\rm rad}^2 & = & {\nu_{\rm K}^2}\,\left(1-6\,x^{-1}+ 8 \,a \, x^{-3/2} -3 \, a^2 \, x^{-2} \right), \\
\nu_{\rm vert}^2 & = & {\nu_{\rm K}^2}
\,\left(1-4\,a\,x^{-3/2}+3a^2\,x^{-2}\right), \nonumber
\end{eqnarray}
\noindent where $x = r/(GM/c^2)$ is the dimensionless radius expressed in terms of the gravitational radius of the black hole. 

For a particular resonance n:m, the equation

\begin{equation}
\mathrm{n}\nu_{\rm rad} = \mathrm{m}\nu_{\rm}; \quad \nu=\nu_\theta\,~\mathrm{or}~\,\nu_\mathrm{K}
\end{equation}
determines the dimensionless resonance radius $x_{\mathrm{n}:\mathrm{m}}$ as a function of spin $a$
 (see Fig. \ref{Figure2}) \footnote{Because of the properties of Kerr black hole spacetimes, \emph{any} relativistic model of black hole QPOs should be rather sensitive to the spin $a$, however this sensitivity can be negligible on large scales of mass (\citealt[]{AbramowiczEtal2004a}) -- see Fig. \ref{Figure5}.}.
Thus, from the observed frequencies and from the estimated mass one can calculate the relevant spin of a central black hole (\citealt[]{AbramowiczKluzniak2001,TAKS}).

\subsection{Microquasars observational data}

\begin{figure*}[t!]
\centering
\begin{minipage} {.48\hsize}
\includegraphics[angle=-90, width=\hsize]{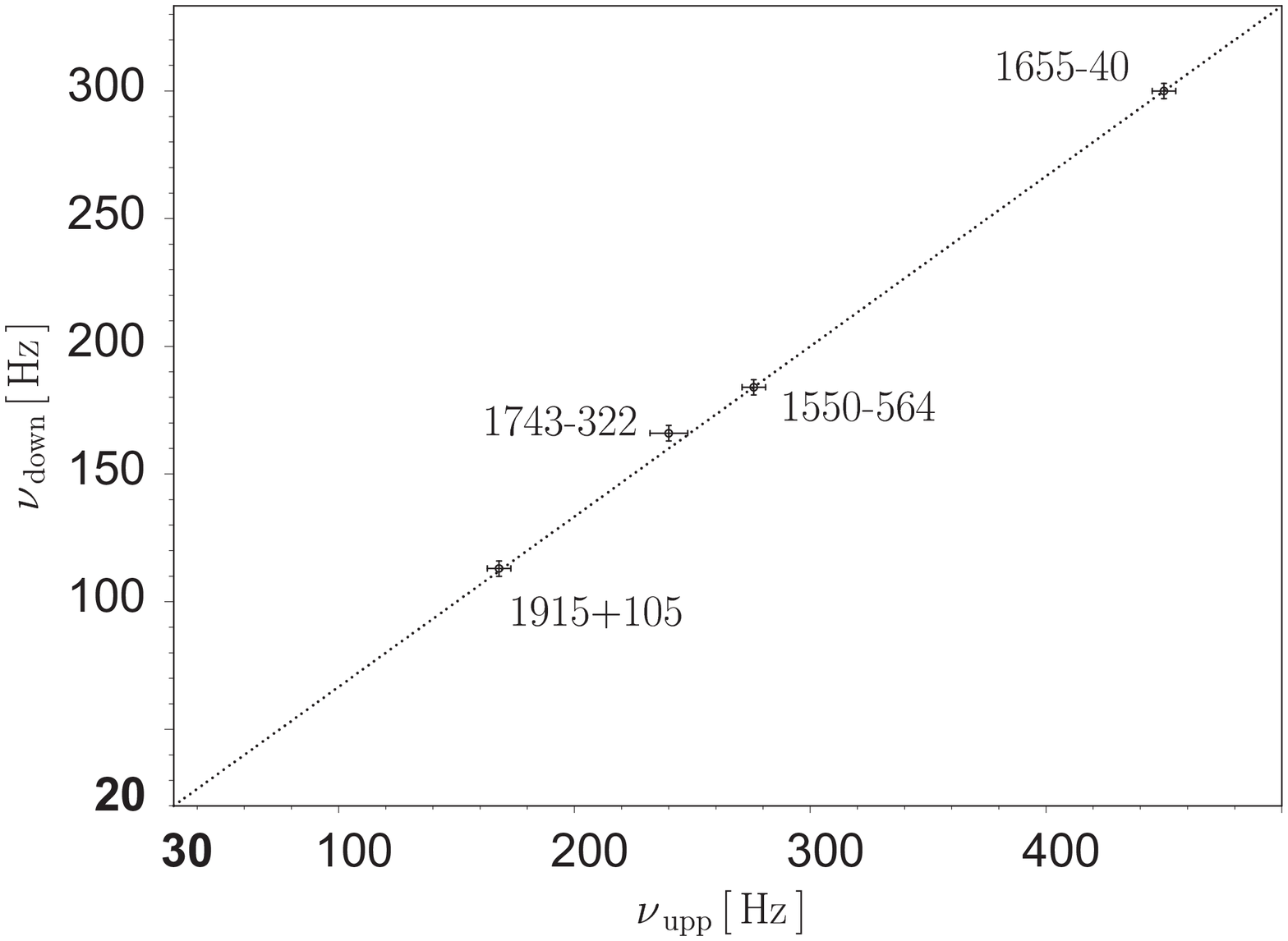}
\end{minipage}
\hfill
\begin{minipage} {.48\hsize}
\includegraphics[angle=-90, width=\hsize]{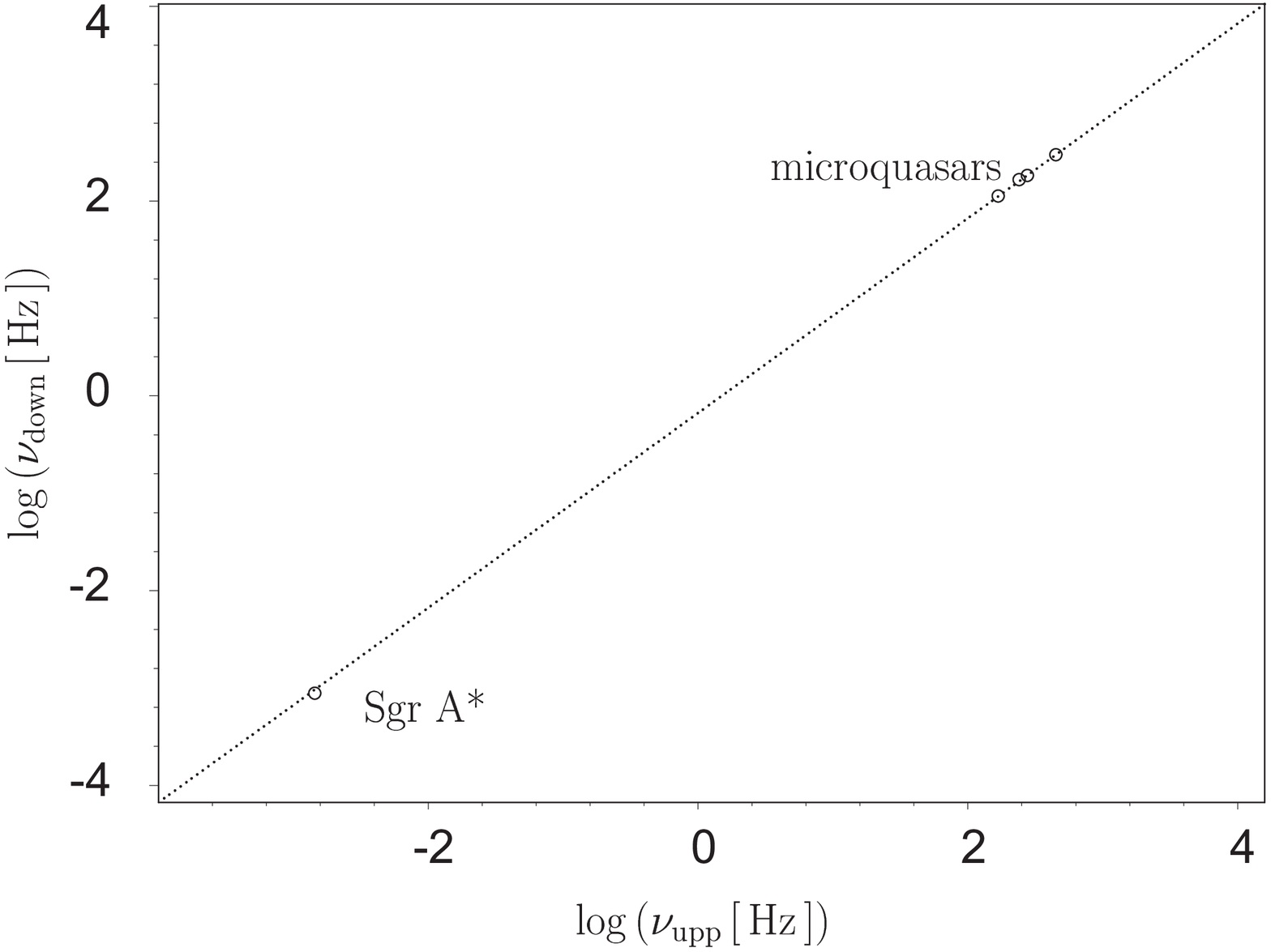}
\end{minipage}
\caption{\label{Figure3}{\it Left:} In all four microquasars where double peak kHz QPOs were detected, the observed frequencies $\nu_{\rm upp}$ and $\nu_{\rm low}$ are clearly in the 3:2 ratio.
{\it Right:} The same 3:2 ratio seems to be present in double peak QPOs in Sgr A*.
}
\end{figure*}

Till this time, double peak kHz QPOs were measured in the case of four microquasars (GRO 1655--40; GRS 1915+105, XTE 1550--564, H 1743--322) and in every case the data show 3:2 ratio of frequencies in the double peak ($\nu_\mathrm{upp}/\nu_\mathrm{down}=3/2$). Mass estimates for central black holes are known and moreless firmly established for the three of these four sources, however for the GRS 1915+105 the mass estimate still varies of factor about two (\citealt[]{Greiner2001}) and for GRO 1655--40 exists two incompatible studies of the mass (\citealt[]{Greene2001,Beer2002}) -- for the summary of microquasars data see Table \ref{Table1} together with its references, while Fig. \ref{Figure3} shows the illustration of significant 3:2 ratio.

\begin{table*}[t!]
\begin{center}
      \caption[]{\label{Table1}Frequencies of twin peak QPOs in microquasars and Galaxy centre black hole}
\renewcommand{\arraystretch}{1.2}
\begin{tabular}{ l  c  c  c  c  c  l }
            \hline
    Source$^{~\rm (a)\,}$ &  $\nu_{\rm{upp}}\,$[Hz]&$\Delta\nu_{\mathrm{upp}}\,$[Hz]& $\nu_{\rm {down}}\,$[Hz]&$\Delta\nu_{\rm{down}}\,$[Hz]&$ {2\nu_{\rm{upp}}/3\nu_{\rm{down}}- 1}$& Mass$^{~\rm (b)\,}$ [\,M$_{\odot}$\,] \\
\hline
\hline
GRO~1655--40    & 450&$\pm\,3$& 300&$\pm\,5$ &\phantom{-}0.00000& $\phantom{1}$6.0\, --- $\phantom{1}$6.6   \\
XTE~1550--564   & 276&$\pm\,3$& 184&$\pm\,5$& \phantom{-}0.00000& $\phantom{1}$8.4\, --- 10.8   \\          
\phantom{RS}H~1743--322   & 240&$\pm\,3$& 166&$\pm\,8$&-0.03614& not measured    \\
GRS~1915+105    & 168&$\pm\,3$& 113&$\pm\,5$& \phantom{-}0.00885& 10.0 \,--- 18.0   \\
\hline
Sgr~A*\phantom{743--322}   & 1.445&$\pm 0.16$ mHz  &0.886&$\pm 0.04$ mHz  & \phantom{-}0.08728&(2.6 --- 4.4)\,$\times10^6$\\
            \hline
        \end{tabular}
\begin{list}{}{}
\item[$^{\rm{(a)}}$] Twin peak QPOs first reported by \citet{Strohmayer2001,RemillardMunoMcClintockOrosz2002,Homanetal2003,RemillardMunoMcClintockOrosz2003,Aschenbach2004a}.
\item[$^{\rm{(b)}}$] See \citet{Greene2001,Orosz2002,Greiner2001,McClintockRemillard2003} for the microquasars. Note that there is the different estimate for GRO~1655--40: $M=(5.4 \pm 0.3)\, {\rm M}_\odot$ (\citet[]{Beer2002}). 
\\
Interval for Sgr$\,$A$^*$ (used in \citealt[]{TAKS}) is resulting from several recent analysis.
\end{list}
\end{center}
\end{table*}

\subsection{The Sgr$\,$A$^*$ data}

The mass of Galaxy centre black hole have been discussed in several recent studies and the best estimate is usually given as $(3.6 \pm 0.4) \times 10^6~{\rm M}_{\odot}$ (e.g. \citealt[]{Schoedeletal2002,Schoedeletal2003,Eisenhaueretal2003,Ghezetal2003,Ghezetal2004}); however substantially lower mass is not excluded yet (\citealt{Reidetal1999,Reidetal2003,BackerandSramek1999,Schoedeletal2003}).

\citet{Aschenbach2004a} have reported five QPOs periodicities in the Galaxy centre black hole X--ray variability: 692 sec, 1130 sec, 2178 sec, 100 sec and 219 sec. Shortly before the Aschenbach's measurement, \citet{Genzeletal2003} found a clear periodicity of 17 min (1020 sec) in Sgr A* infra-red variability during a flaring event.
It is rather difficult to confirm the Aschenbach's data, nevertheless note that 
(\citealt[]{AbramowiczEtal2004a,AbramowiczEtal2004b,Aschenbach2004b, TAKS}):
$$(1/692):(1/1130):(1/2178) \approx 3:2:1.$$

\begin{figure*}[t!]
\centering
\begin{minipage} {.46\hsize}
\medskip

\includegraphics[angle=-90, width=\hsize]{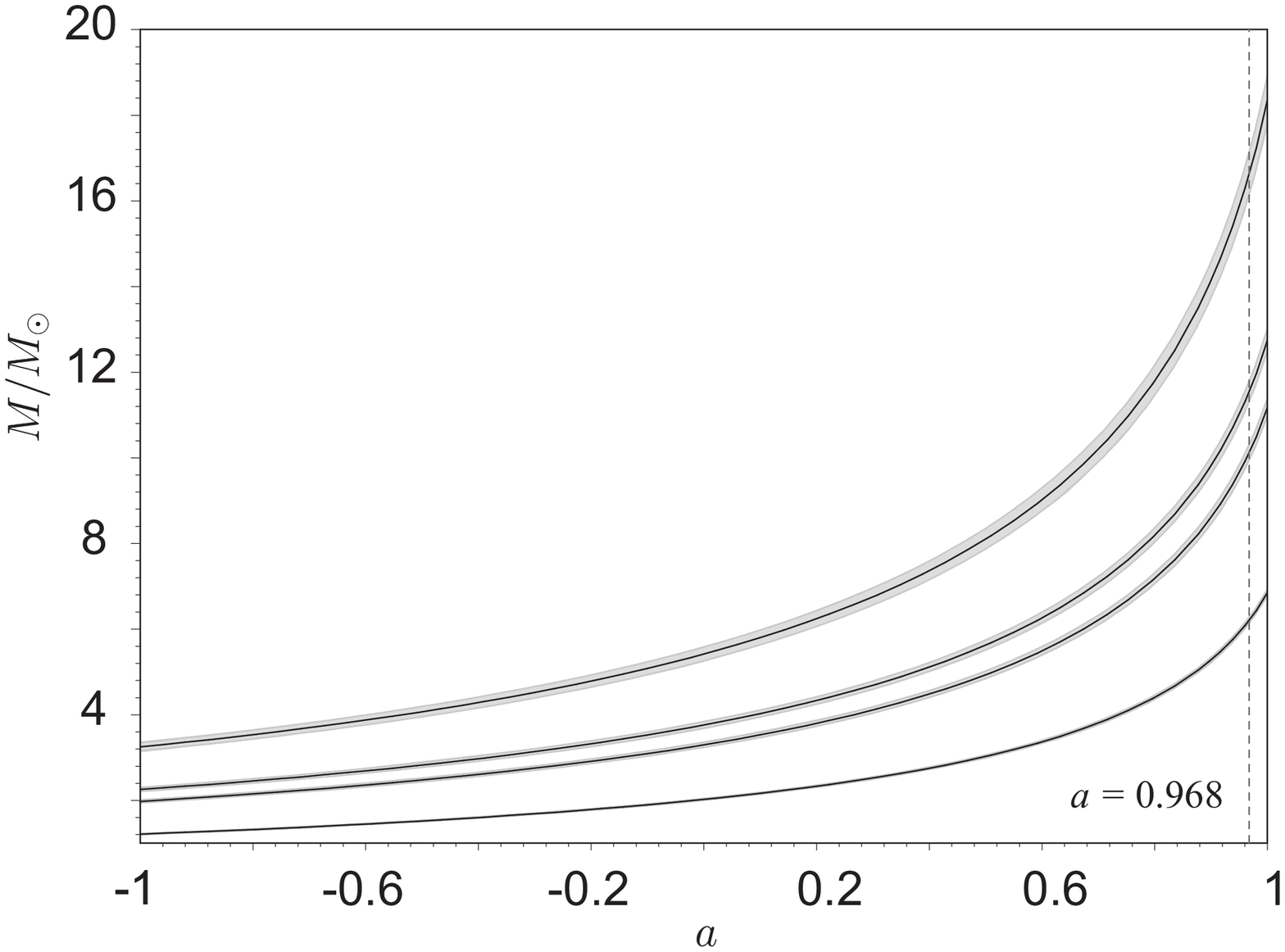}
\end{minipage}
\hfill
\begin{minipage} {.5\hsize}
\includegraphics[angle=-90, width=\hsize]{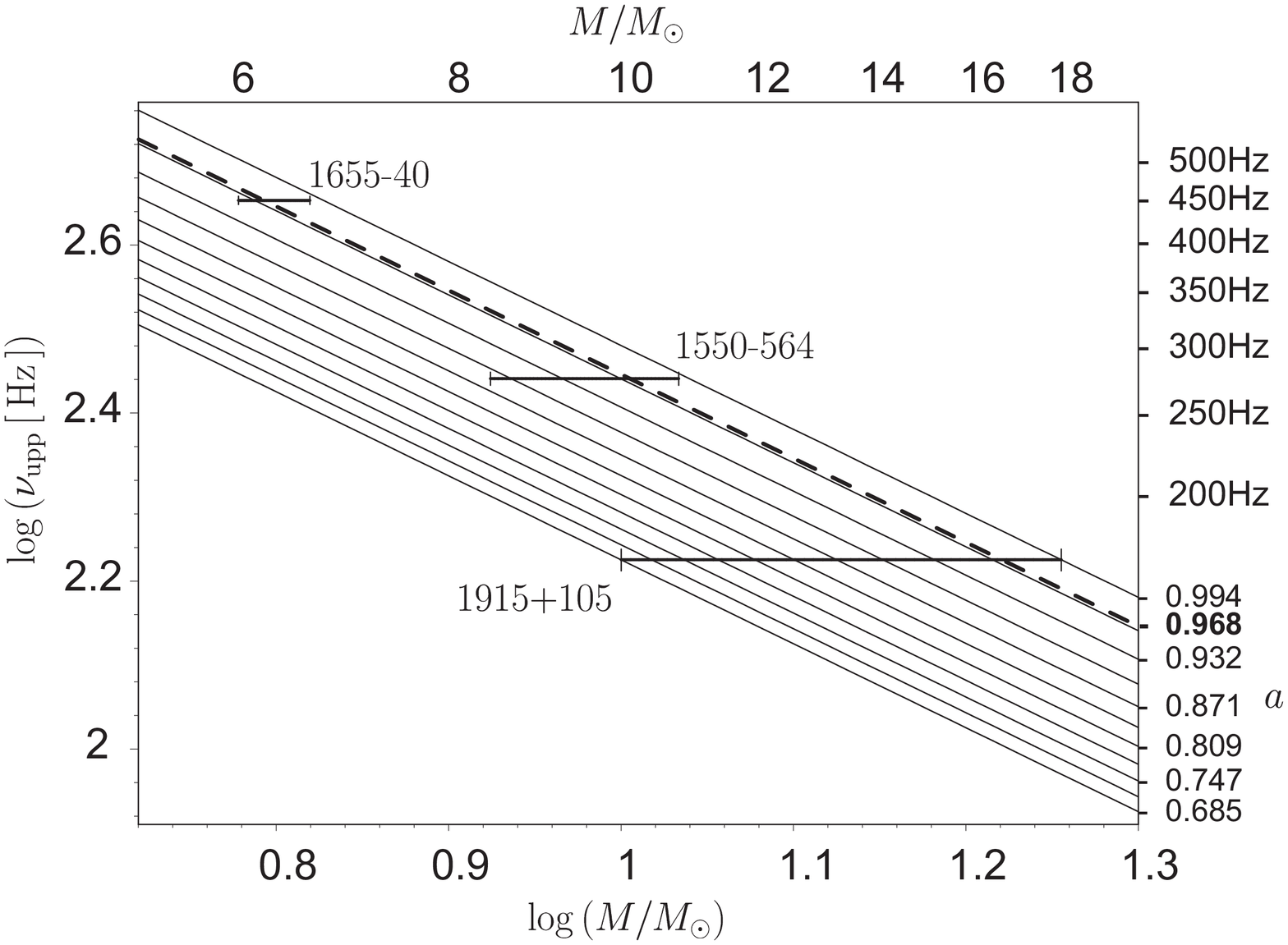}
\end{minipage}
\caption{\label{Figure4} {\it Left:} Mass-spin dependence $M(a)$ calculated from the 3:2 parametric resonance model for the frequencies $\nu_\mathrm{upp}=$ 168 Hz, 242 Hz, 276 Hz, and 450 Hz (from the top in this order) observed in four microquasars. Shadows show the range $\pm5\,$Hz. {\it Right:} the same in the form of fit to the observational data. For both panels, the dashed line $a=0.968$ corresponds to the
observational fit $\nu_\mathrm{upp}=2.793\ {\rm M}_\odot/M\,\mathrm{kHz}$ found by \cite{McClintockRemillard2003} .
}

\end{figure*}
\begin{figure*}[t!]
\centering
\begin{minipage} {.5\hsize}
\bigskip\smallskip

\includegraphics[angle=-90, width=\hsize]{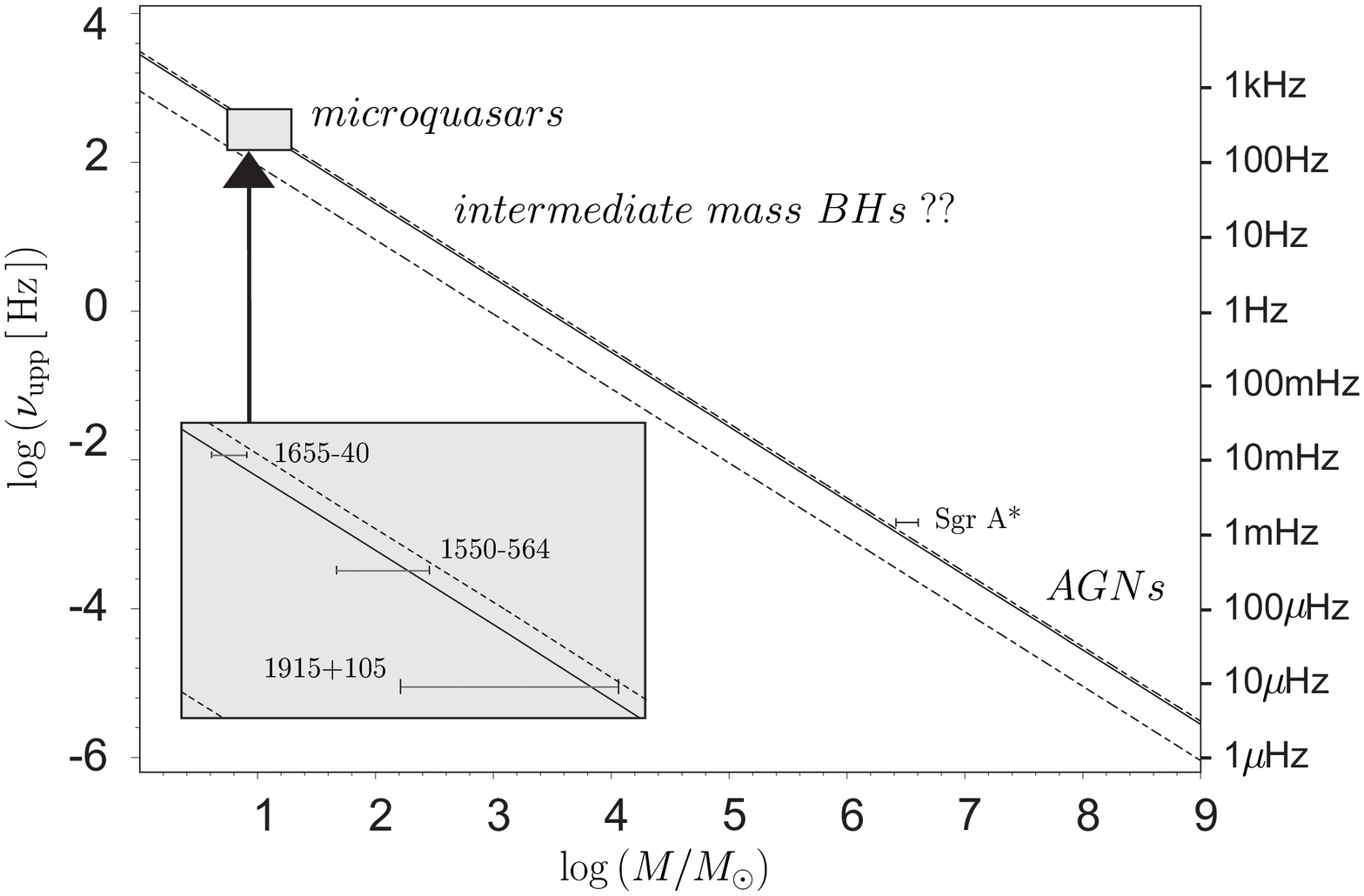}
\end{minipage}
\hfill
\begin{minipage} {.46\hsize}
\includegraphics[angle=-90, width=\hsize]{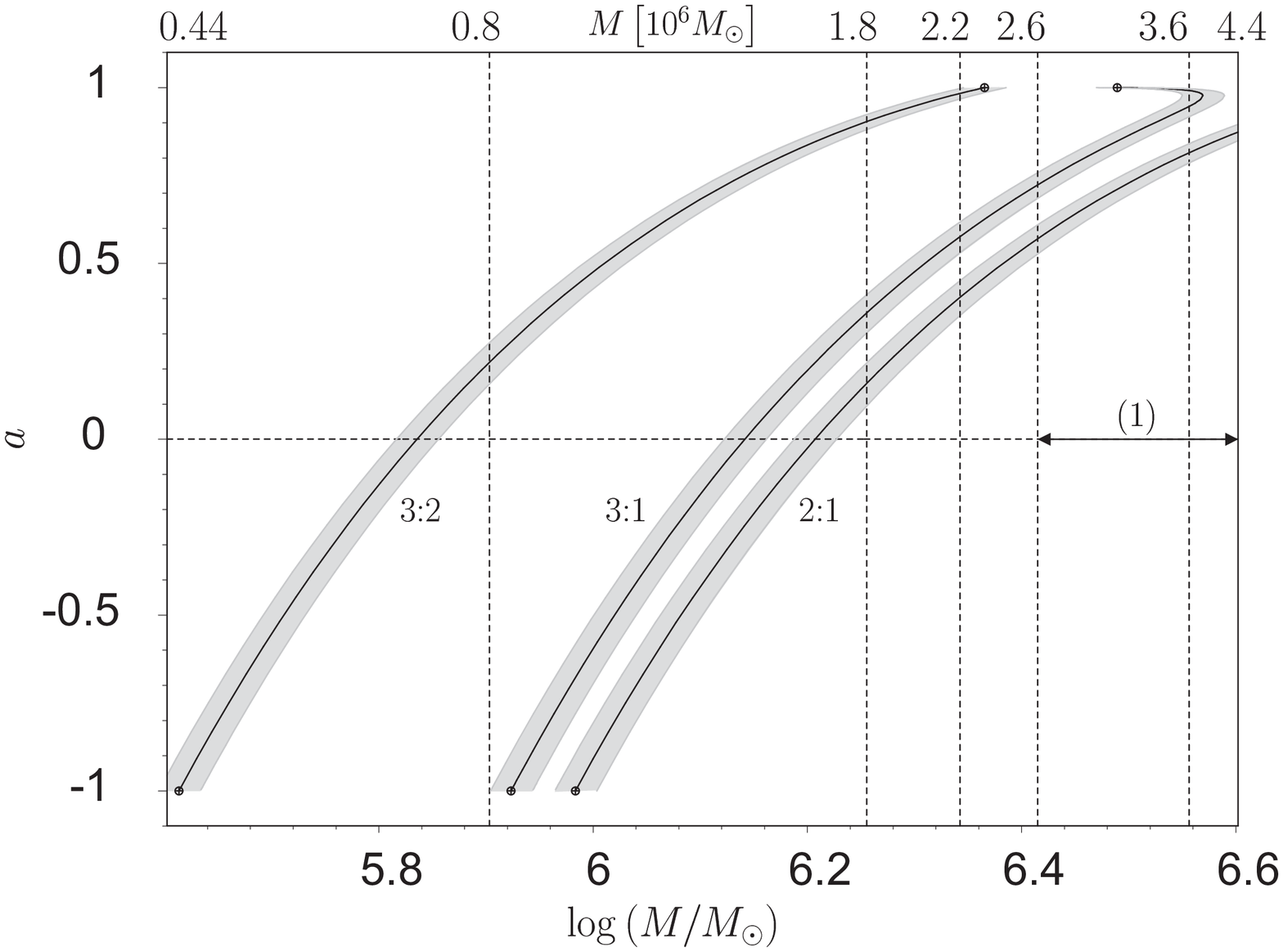}
\end{minipage}
\caption{\label{Figure5}
{\it Left:} 
The upper frequency of the double peak predicted as the function of mass by 3:2 parametric resonance model. Mass range for Sgr A* corresponds to the range given in Table \ref{table1}. Dotted lines are for $a=0\,$(lower line) and  $a=1\,$(upper line), the solid line is the fit $1/M$ found by \citet{McClintockRemillard2003}.
{\it Right:} 
Spin dependence for 3:2 parametric, 3:1 and 2:1 forced resonance in Sgr A* implied by $\nu_\mathrm{down}=0.886\,\mathrm{mHz}$ measured by \citet{Aschenbach2004a}, shadows respect accuracy states in that measurement. For the 3:1 forced resonance and very high spin, the estimate is ambigous what is valid for any epicyclic resonance with $p/q>2.18$ - see \citet{TS} where the properties of epicyclic frequencies in the Kerr spacetimes are discussed in detail.
}
\end{figure*}

\subsection{Estimates of the spin from particular models (quantitative results)}

Angular momentum estimates are shown in Table \ref{Table3} for several particular resonance models, while Fig. \ref{Figure4} illustrates how the 3:2 parametric resonance model fits the data. For the Sgr$\,$A$^*$, we present the estimate of spin in
Fig. \ref{Figure5} (right panel).

\section{Discussion and conclusions}

In the case of microquasars all resonances discussed here (except the 3:2 Keplerian resonance) are consistent with the existing microquasars data and comparing this data with predictions of the theory gives a clear estimate of the spin for given source and model. The different relativistic QPOs models can be compared at the moment when some independent and convincing estimate of the microquasars spin will be done. For example, the 3:2 parametric resonance model
\footnote{One should remind that, in difference to forced resonances, the 3:2 parametric resonance model gives the 3:2 observed ratio naturally without any additional assumptions.} 
expects high spin ($a\sim0.95$) for the microquasars, while another {relativistic precession model} of \citet{Stella} predicts the spin to be rather low ($a\sim0.2$).
Unfortunately, no such clear spin predictions exist at the present stage. For example, it is often argued that the presence of relativistic jet (which the discussed microquasars show) is a signature of large black hole spin (\citealt[]{BlandfordZnajek1977}) but there is some evidence against (\citealt[]{GhoshAbramowicz1997}). Also the up-and-coming studies of {spectral iron lines} give contradictory results as well (see, e.g., \citealt[]{Karas2002}).

In the case of Galaxy centre black hole the spin estimate following from the measurement of the 3:2 QPOs frequencies (1/692):(1/1130) and the most accepted prediction $M_{\mathrm{Sgr}\,\mathrm{A}^{*}}\sim3.6\times10^{6}\,{\rm M}_\odot$ rather exlude the 3:2 parametric resonance and imply moderate black hole internal angular momentum for the eventual 2:1 forced resonance while for the 3:1 resonance the spin would be rather high. But, in fact, it is difficult to discuss this spin predictions seriously --  the mass of Sgr$\,$A$^*$ is known with the accuracy of one order of magnitude in present and the frequency measurement by \cite{Aschenbach2004a} is not confirmed yet. So one can expect that in principle no of the resonances considered here is excluded (see Fig. \ref{Figure5}).

Finally we stress that the 3:2 ratio seems to be now significant for the compact X-ray sources
\footnote{For black holes see references in the Table \ref{Table1} and \citet{KlisNordita}. For neutron stars see, e.g,  \citet{AbramowiczBursa,Belloni2005}, or \citet{BulikNordita,BarretNordita}.
Overview of the data and detailed references are given in \citet{KlisReview}.}
and the 3:2 data of \cite{Aschenbach2004a} strongly supports the idea of relativistic scaling for QPOs frequencies \citep{AbramowiczEtal2004a} -- if confirmed, these would be of a fundamental importance for the black hole accretion theory and the precise measurement can help to solve the question of QPOs nature.

\newlength{\sirkaA}\newlength{\sirkaB}\settowidth{\sirkaA}{+}\settowidth{\sirkaB}{-}\advance\sirkaA by -\sirkaB
\newcommand{\csp}{$\hspace{\sirkaA}$}
\begin{table*}[t!]
\begin{center}      
\caption[]{\label{Table3}Summary of angular momentum estimates from resonance models for microquasars}
\begin{tabular}{ l  l  l  l  l }    
\hline
~& \multicolumn{4}{c}{Interval of possible spin $a$ relevant for}\\\\
Model for &~~~~~1550--564 &~~~~~1655--40 &~~~~~1655--40$^*$ &~~~~~1915+105\\
\hline
\hline
3:2 [$\nu_{\theta},~\nu_r$] ~~ \phantom{``}parametric &
+0.89~---~+0.99 &+0.96~---~+0.99 &+0.88---~+0.93 &+0.69~---~+0.99\\  
\hline
2:1 [$\nu_{\theta},~\nu_r$] ~~ \phantom{``}forced &+0.12~---~+0.42 &+0.31~---~+0.42 &+0.10---~0.25 & \csp -0.41~---~+0.44\\ 
\hline
3:1 [$\nu_{\theta},~\nu_r$] ~~ \phantom{``}forced & +0.32~---~+0.59 & +0.50~---~+0.59 &+0.31---~+0.44 & \csp-0.15~---~+0.61\\ 
\hline
\hline
3:2 [$\nu_{\rm K},~\nu_r$] ~~``Keplerian'' p. & $\phantom{+0.00}$~$\phantom{---}$~$\phantom{+0.00}$ & $\phantom{+0.00}$~$\phantom{---}$~$\phantom{+0.00}$ & $\phantom{+0.00}$~$\phantom{---}$~$\phantom{+0.00}$ & +0.79~$\phantom{---}$~$\phantom{+0.00}$\\ 
\hline
2:1 [$\nu_{\rm K},~\nu_r$] ~~``Keplerian'' f.& +0.12~---~+0.43 & +0.31~---~+0.42 &+0.10---~+0.25 & \csp-0.41~---~+0.44\\ 
3:1 [$\nu_{\rm K},~\nu_r$] ~~``Keplerian'' f.& +0.29~---~+0.54 & +0.45~---~+0.53 &+0.28---~0.40 & \csp-0.13~---~+0.55\\          
     \hline
     \end{tabular}
     \end{center}
 {Table contains values of dimensionless spin calculated exactly from the upper observed frequency, error $\Delta a$ resulting from uncertainty of frequency measurement $\Delta \nu_\mathrm{upp}$ is for microquasars XTE 1550-564 (GRO 1655-40, GRS 1915+105) given as $\sim\pm\,0.03$ (0.01, 0.05).
\smallskip 
 
$^*$ This second column for GRO 1655--40 shows numbers outgoing from the mass analysis by \cite{Beer2002}. Theirs estimate $M_{~\mathrm{GRO}\,1655-40}\sim 5.4\,{\rm M}_\odot$ is nearly one solar mass lower than that by \cite{Greene2001}. Note that while the estimate of spin from 3:2 parametric resonance is for both cases ($6.3;\,5.4~M_\odot$) high and in fact similar: $a\approx 0.9$, for other discussed eventualities this difference in the mass results in large difference of spin.}
\end{table*}

\acknowledgements

I thank Marek Abramowicz, Wlodek Kluzniak and Zdenek Stuchlik for discussion and help. This work was supported by the Czech grant MSM 4781305903. I also thank the excellent hospitality of Nordita (Copenhagen).

\end{document}